\documentclass[conference]{IEEEtran} 

\usepackage[T1]{fontenc} 
\usepackage{amsmath,amssymb,physics} 
\usepackage{algorithm}
\usepackage{algpseudocode}
\usepackage{float}
\usepackage{graphicx} 
\usepackage{cite} 
\usepackage{balance} 
\usepackage{qcircuit}
\usepackage{subcaption}
\usepackage{todonotes}
\usepackage{comment}
\usepackage{siunitx}
\usepackage{booktabs}
\usepackage{blindtext}
\usepackage{svg}
\usepackage{url}
\usepackage{xurl}

\usepackage{array}
\usepackage{balance} 
\usepackage{blindtext}
\usepackage{bm}
\usepackage{booktabs}
\usepackage{comment}
\usepackage{datetime}
\usepackage{fancyhdr}
\usepackage{float}
\usepackage{graphicx} 
\RequirePackage[colorlinks=true, allcolors=blue]{hyperref}
\usepackage{makecell}
\usepackage{multirow}
\usepackage{mwe}
\usepackage{siunitx}
\usepackage{subcaption}
\usepackage{svg}
\usepackage{tablefootnote}
\usepackage[absolute]{textpos}
\usepackage{url}
\usepackage{xcolor}
\usepackage{xurl}

\definecolor{ionqorange}{HTML}{FF5000}
\hypersetup{
    colorlinks=true,
    linkcolor=ionqorange,    
    citecolor=ionqorange,    
    urlcolor=ionqorange      
}

\IEEEoverridecommandlockouts 

\begin{document}

\title{Measuring Accuracy and Energy-to-Solution of Quantum Fine-Tuning of Foundational AI Models}
\author{
    \IEEEauthorblockN{
        Oliver Knitter\IEEEauthorrefmark{1}\IEEEauthorrefmark{2},
        Sang Hyub Kim\IEEEauthorrefmark{1}\IEEEauthorrefmark{2},
        Maximilian Wurzer\IEEEauthorrefmark{4},
        Jonathan Mei\IEEEauthorrefmark{2},
        Claudio Girotto\IEEEauthorrefmark{2},\\
        Karen Horovitz\IEEEauthorrefmark{2},
        Chi Chen\IEEEauthorrefmark{2},
        Masako Yamada\IEEEauthorrefmark{2},
        Frederik F. Fl\"{o}ther\IEEEauthorrefmark{4}\IEEEauthorrefmark{3},
        Martin Roetteler\IEEEauthorrefmark{2}
        \medskip
    }
    \IEEEauthorblockA{
        \IEEEauthorrefmark{2}IonQ Inc., 4505 Campus Dr, College Park, MD 20740, USA\\ \{oliver.knitter, sang, jonathan.mei, claudio.girotto, karen.horovitz, chi.chen,\\ yamada, martin.roetteler\}@ionq.co}
    \IEEEauthorblockA{    \IEEEauthorrefmark{4}QuantumBasel, Schorenweg 44b, 4144 Arlesheim, Switzerland\\ \{maximilian.wurzer, frederik.floether\}@quantumbasel.com}
    \IEEEauthorblockA{
        \IEEEauthorrefmark{3}Center for Quantum Computing and Quantum Coherence (QC2),
University of Basel,\\ Klingelbergstrasse 82, 4056 Basel, Switzerland}
}

\maketitle

\begingroup
\renewcommand\thefootnote{\IEEEauthorrefmark{1}}

\makeatletter\def\Hy@Warning#1{}\makeatother
\footnotetext{Equal contribution}
\endgroup

\begin{abstract}
We present an experimental study of energy-to-solution (ETS) of hybrid quantum-classical applications, enabled by direct instrumentation of power consumption of a Forte Enterprise trapped-ion quantum processor. We apply this methodology to a hybrid quantum-classical pipeline for quantum fine-tuning of foundational AI models, and validate the approach end-to-end on quantum hardware. Despite noise and limited qubit counts, the resulting models achieve accuracy competitive with and exceeding classical baselines such as logistic regression and support vector classifiers. Our results show that QPU energy consumption scales approximately linearly with qubit number for shallow circuits, while classical simulation exhibits exponential scaling, indicating a break-even for ETS around 34 qubits. 
The classification error improvement of the best quantum fine-tuned model over the best classical fine-tuned model considered in this study is around 24\%.
We further contextualize these findings with comparisons to tensor network methods.
This work establishes energy-to-solution as a measurable and scalable metric for evaluating quantum applications and provides experimental evidence of favorable energy--accuracy trade-offs.

\end{abstract}

\section{Introduction}
While a considerable amount of technological progress still remains before quantum computers see mainstream commercial adoption, the last few years have seen a rapid progression in the development of quantum hardware and algorithms. Though computational speed, so far only accessible by fault-tolerant quantum hardware, remains the most analyzed source of quantum advantage~\cite{aaronson2022much}, the field of quantum machine learning (QML) has begun exploring other potential sources, such as accuracy~\cite{peral2024systematic}, model size and memory requirements~\cite{zhao2026exponential}, energy efficiency~\cite{jaschke2023quantum}, and the ability to handle high-dimensional, noisy, or otherwise difficult-to-analyze datasets~\cite{peters2021machine, marshall2023high}. Algorithms excelling in these areas might already help bring some forms of quantum advantage in the current era of noisy intermediate-scale quantum (NISQ) hardware.

At the same time, the steady development of powerful large language models (LLMs) and foundation models has precipitated great strides in the field of artificial intelligence (AI). As these grow increasingly bigger, with the largest containing hundreds of billions of trainable parameters, some of the ongoing public discourse surrounding LLMs has centered around the considerable amounts of energy they require, with recent estimates placing the energy cost of training a single LLM at around 1.5 terajoules~\cite{luccioni2023estimating}. In response, techniques like test-time training~\cite{sun2407learning} and model fine-tuning~\cite{jung_joint_2015,kading_fine-tuning_2017, wang_growing_2017} help mitigate energy costs by providing viable alternatives to retraining entire LLMs from scratch.

A natural hypothesis to consider is the potential that exists, either now or in the future, for quantum computers to improve AI energy efficiency. This paper expands on prior work demonstrating the utility of trainable quantum circuits as final layer fine-tuning heads for LLMs~\cite{kim2025quantum} by experimentally validating previous estimates for the energy needed to perform inference runs of these fine-tuning circuits on a state of the art trapped-ion quantum processor. We also present the addition of several circuit design changes, error mitigation, and classical postprocessing techniques to the base architecture from~\cite{kim2025quantum}, which are needed to overcome the inherent noisiness that comes with running these circuits on actual hardware. To the extent of our knowledge, hardware experiments of this scale, designed to analyze the energy cost of performing specific QML tasks, have not yet been presented in the literature: previous studies typically estimate energy consumption by simply multiplying the measured computation time by device-specific average power draw~\cite{berger2021quantum,desdentado2024exploring}. Our experimental findings appear to corroborate this general approach. More specifically, we find that for hardware-efficient constant-depth circuits, execution time and energy consumption are both roughly linear with respect to qubit number.

Prior work related to this area includes comparisons made between quantum and classical, specifically tensor network, algorithms~\cite{fu2024achieving}, analyses of the fundamental cost to effectively cool quantum systems~\cite{taranto2023landauer,taranto2025efficiently}, and the broader impact of quantum technology on general energy consumption~\cite{auffeves2022quantum}. It has been proven, at least for some calculations, that the computational efficiency gains of quantum computers over their classical counterparts can translate into commensurate gains in energy efficiency~\cite{meier2025energy}. Moreover, though the relevant metrics are still being debated~\cite{jaschke2023quantum,chen2023quantum}, full-stack quantum computers can themselves be optimized to yield improvements in resource and energy efficiency~\cite{fellous2023optimizing}.

In Section \ref{sec:preliminaries}, we recap the language model fine-tuning setup from~\cite{kim2025quantum}. Section \ref{sec:methodology} describes the changes to model architecture and training that we have implemented in this paper, specifically concerning noise mitigation strategies used to improve performance when running these circuits on an IonQ Forte Enterprise system. Section \ref{sec:results} discusses our data analysis and findings, while we devote some time in Section \ref{sec:mps_comparison} toward discussing the implications of our results when comparing the QPU's energy consumption to classically tractable simulation via matrix product states (MPS). In Section \ref{sec:conclusion} we summarize our work and provide guidance for future research. Altogether, the main findings of this paper are as follows:
\begin{itemize}
    \item For depth-limited hardware-efficient circuits run on a QPU, both computational time and energy consumption are roughly linear with respect to qubit number, in contrast with classical statevector simulations.
    \item There does not appear to be a strong correlation between the average power draw of a QPU and the size (width and depth) of the circuits it runs.
    \item Commonplace error mitigation techniques are effective for improving circuit performance on NISQ hardware.
    \item It is reasonable to extrapolate from our findings that NISQ QPUs are likely both time- and energy-efficient at scale relative to MPS, for a class of circuit ans\"atze that is computationally viable for both quantum and classical backends.
\end{itemize}

\section{Preliminaries}
\label{sec:preliminaries}

This section provides a high-level discussion of the workflow introduced in~\cite{kim2025quantum}, which our study follows, for training quantum circuits to fine-tune classical language models. Fine-tuning is a form of transfer learning that involves modifying a preexisting model, either by further updating its learned weights or by training newly introduced model components while keeping the original weights frozen, in order to hone the model's performance on a small dataset representing a task for which the base model was not trained to perform. Considerably less computationally intensive than training a new model from scratch, fine-tuning itself thus serves as a potential avenue for mitigating additional growth of the energy footprint associated with AI model training. We now discuss the relevant background for how a quantum circuit can be used to fine-tune an otherwise classical model.

\subsection{Sentence transformer and SetFit}


Following~\cite{kim2025quantum}, the experiments conducted in this study comprise the specific task of fine-tuning a sentence transformer~\cite{reimers_sentence-bert_2019}, a BERT-based~\cite{devlin_bert_2019} encoder-only language model designed to convert entire sequences of text into latent embedding vectors. SetFit~\cite{tunstall_efficient_2022} is a training framework for fine-tuning sentence transformers, using contrastive learning to train a separate neural network head operating on the sentence transformer's latent vector outputs. The existing pipeline is well amenable to incorporation of QML: the quantum fine-tuning framework simply replaces the classical fine-tuning head with a parameterized quantum circuit. Moreover, SetFit is intended to produce effective training on few-shot text classification tasks, a low-data regime for which QML techniques are potentially well-suited. Following the basic experiments in~\cite{kim2025quantum}, we perform our fine-tuning analysis using the Stanford Sentiment Treebank (SST2)~\cite{socher_recursive_2013} benchmark, a labeled dataset of positive and negative movie reviews.

\subsection{QML and language model fine-tuning}

We assume a baseline familiarity with Dirac notation and the basics of gate-based quantum computing, for which~\cite{nielsen_quantum_2002} provides the quintessential treatment. QML is the broad subfield of quantum computing focused on performing typical machine learning (ML) tasks with quantum computers. Though QML mainly concerns itself with quantum primitives that replace specific subroutines commonly found in traditional ML algorithms, it more generally comprises full algorithms, simulation environments, and models corresponding with significant proportions of classical training and inference pipelines.

\begin{figure*}[hbt]
    \centering\includegraphics[width=1\linewidth]{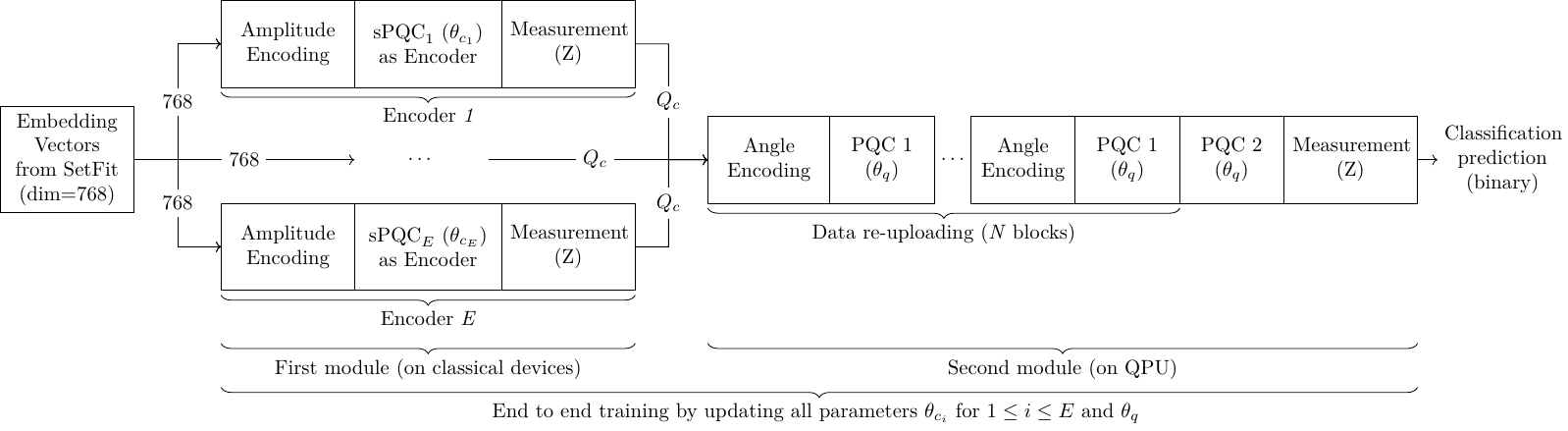}
    \caption{Diagram of the fully-differentiable QML classification head added to the base LLM, reproduced and modified from~\cite{kim2025quantum}. It is composed of two sub-modules: a quantum-inspired multi-head encoder to classically reduce the dimension of the SetFit latent vectors, and a data-reuploading parameterized quantum ansatz operating on a QPU. Note that in contrast to the classification head used in~\cite{kim2025quantum}, our variation of this architecture does not incorporate a final classical linear layer.}
    \label{fig:full_block_diagram}
\end{figure*}

The authors of \cite{kim2025quantum} argue that the capability of QML models to surpass practically achievable classical performance lies, as shown through the correspondence established not only between certain types of QML models and kernel machines~\cite{schuld_supervised_2021}, but also between kernel machines and deep neural networks~\cite{lee_wide_2019, atanasov_neural_2021}, in the careful selection of both the circuit ansatz and encoding method comprising the model. As depicted in Figure~\ref{fig:full_block_diagram}, the original language model fine-tuning architecture from~\cite{kim2025quantum} incorporates both quantum-inspired and truly quantum model components. The first component, a multi-head quantum-inspired encoder, performs statevector simulations modeling the amplitude encoding, execution, and $Z$-basis expectation value measurements of multiple parallel quantum circuits. This encoder reduces the dimension of the input embedding vectors to one more manageable on NISQ hardware.

The outputs of the simulated quantum circuit are then encoded into the second component, a parameterized quantum circuit directly executed on actual quantum hardware. The measurement outcomes of this component then provide the logits used for classification. The quantum head processes its input data through angle encoding~\cite{schuld_supervised_2021}, by which each input feature is used as the angle of a $Y$-axis Bloch sphere rotation acting on a single qubit:
\begin{equation}
\ket{\psi(\mathbf{x})} = R_Y(x_1) \otimes R_Y(x_2) \otimes \dots \otimes R_Y(x_n) \ket{0}^{\otimes n}.
\end{equation} At the expense of needing to operate on a separate qubit for every input feature dimension, when applied through a re-uploading scheme~\cite{perez-salinas_data_2020} that repeatedly uploads the same angles throughout the circuit, angle encoding produces greatly expressive circuit ans\"atze.



\section{Methodology and Experimental Setup}
\label{sec:methodology}

To validate prior theoretical analysis~\cite{kim2025quantum}, we focus on the quantum component of the hybrid architecture, specifically running the quantum circuit component of the architecture on both QPU and GPU, and analyzing power and energy measurements taken from the two hardware platforms as qubit number increases. Simultaneously, we conduct a separate set of accuracy experiments to demonstrate empirically how the QPU achieves comparable classification performance to noiseless simulation when augmented with error mitigation techniques. In this section, we first describe the modifications made to the model architecture relative to~\cite{kim2025quantum}, then discuss the error mitigation pipeline, and finally detail the hardware, data, and hyperparameter configurations for both the energy and accuracy studies.

\subsection{QML Model Modifications}
\label{sec:model_modifications}

Following a suggestion from~\cite{kim2025quantum}, we do \textit{not} employ a final classical linear layer after the quantum measurement. In the original design, this layer maps multi-qubit expectation values to class logits. When trained in simulation, this layer's weights optimize for the noise distribution of the simulator, rather than the actual QPU. If deployed on hardware, the distribution mismatch causes the linear layer to amplify shot noise. Instead, in our modification, we measure the $Z$-basis expectation value $\langle Z \rangle$ of a single target qubit, using it directly as the binary classification logit. Beyond eliminating the learned parameters most sensitive to the simulation--hardware distribution gap, constraining the model output to the range $[-1, 1]$ in this way yields a more robust decision boundary. At the other end of the pipeline, we also replace the simulated PQC encoder with a linear layer encoder for ease of training, a choice also discussed in~\cite{kim2025quantum}.

The parameterized quantum circuit (PQC) follows the data re-uploading ansatz from~\cite{kim2025quantum}, with hyperparameters as summarized in Table~\ref{tab:hyperparam_letters}. The circuit comprises $M = 2$ main blocks, each comprising a data re-uploading sub-block followed by a trainable entangling layer. Within each re-uploading sub-block, the input data is re-encoded $R = 4$ times: each re-upload applies $R_Y$ angle encoding to all qubits, followed by a trainable entangling layer of CNOT gates and parameterized $R_Y$ rotations. After the two main blocks, a final trainable layer ($N = 1$) completes the circuit. The resulting two-qubit gate count scales linearly with qubit number: 35, 42, 49, 56, and 63 two-qubit gates for 10, 12, 14, 16, and 18 qubits, respectively. To mitigate errors introduced by two-qubit gates, our entangling CNOT ladders contain half as many gates as typical and skip over every other qubit (so that a CNOT acts on qubits $i$ and $i+2$), with ladders alternately operating on even- and odd-indexed qubits.

\subsection{Circuit Transpilation and Debiasing}
\label{sec:debiasing}

Running quantum circuits on trapped-ion hardware introduces systematic errors that depend on the physical ion assignment. To reduce these effects, we use debiasing and non-linear filtering (DNL), a symmetrization-based error mitigation procedure similar to that described in~\cite{symm2023}. For each logical circuit, we generate 25 equivalent variants corresponding to different valid assignments of logical qubits to physical ions. These variants implement the same ideal circuit but sample different hardware-dependent error profiles.

Each variant is executed with a fraction of the total shot budget. We note here that this procedure requires a minimum of 500 shots to be practicable~\cite{symm2023}. The measured bitstrings are then remapped back to the original logical qubit ordering, producing 25 histograms for the same logical circuit. Aggregating these histograms averages over mapping-dependent biases that would otherwise affect any single physical realization.

\subsection{Post-Processing: Non-Linear Aggregation and Bias Correction}
\label{sec:dnl}

After completing the symmetrization step, we aggregate the 25 remapped histograms and then apply an unsupervised bias correction to the resulting logits.

\subsubsection{Non-Linear Aggregation Filter}

The default aggregation method is simple averaging, in which the probability of each bitstring is averaged over all 25 variants. We also consider a non-linear aggregation filter that down-weights bitstrings whose large counts are concentrated in only a small number of variants. This is intended to reduce the influence of variant-specific hardware artifacts that survive the symmetrization step.

The filter proceeds as follows:
\begin{enumerate}
    \item For each measured bitstring $b$, collect its observed frequencies $\{f_{vb}\}$ across all $V = 25$ variants and sort them in descending order.
    \item Transpose the sorted frequency distribution to obtain, for each frequency level $f$, the number of variants $v$ that simultaneously observe $b$ with frequency at least $f$.
    \item Apply a power-law filter with exponent $p$: each point on the cumulative curve is weighted by $W(v) = (v / V)^p$. Higher values of $p$ more aggressively suppress bitstrings appearing with high frequency in only a small subset of variants.
    \item A threshold parameter $t$ further discards contributions from bitstrings appearing in fewer than $t$ variants.
    \item The aggregated probability for each bitstring is the normalized area under the filtered curve.
\end{enumerate}
This filter preserves bitstrings that appear consistently across many variants, while also suppressing bitstrings that appear strongly in only a small subset of variants. When $p = 0$ and $t = 0$, the procedure reduces to simple averaging. For each qubit count, we selected the filter parameters $(p,t)$ by grid search over a fixed set of candidate values, using classification accuracy on the evaluation set. We therefore interpret the filtered accuracy as a calibrated upper bound rather than as the baseline hardware result.

\subsubsection{Bias Correction}

Following the histogram aggregation, we compute the expectation value $\langle Z \rangle_i$ of the target qubit for each of the $N$ test samples. The distribution of logits may be shifted from its ideal center by a systematic hardware-induced bias, which we correct by subtracting the global mean logit:
\begin{equation}
    \langle Z \rangle_i^{\text{bc}} = \langle Z \rangle_i - \frac{1}{N} \sum_{j=1}^{N} \langle Z \rangle_j.
\end{equation}
The corrected logits produce the output classification: a sample is assigned to Class 0 if $\langle Z \rangle_i^{\text{bc}} > 0$ and Class 1 otherwise. This correction is unsupervised, requiring no labeled data and resulting solely from the test-time measurements.

\subsection{GPU consumption estimates}

To measure the energy consumption used by executing these circuits on a simulated classical backend, we used the detailed profiling tool CodeCarbon~\cite{benoit_courty_2024_11171501} to estimate and track the energy consumed by both the CPU and GPU. We found CodeCarbon behaved more consistently in an isolated virtual machine (VM) than in a shared Slurm cluster~\cite{yoo2003slurm}, where parts of other unrelated workloads were also visible to the profiler.
Thus, to capture data center conditions while still having some level of control over the runtime environment, we used VMs with attached accelerators in Google Cloud Platform (GCP)\footnote{\url{https://cloud.google.com/compute/docs/accelerator-optimized-machines}}, specifically with the following configuration: a \texttt{g2-standard-16} with 16 threads of a 2.20 GHz Intel Xeon CPU, 64GB RAM, and an NVIDIA L4 24GB~\cite{noauthor_nvidia_nodate_l4} GPU. To run quantum algorithms on GPU, we used PennyLane~\cite{bergholm_pennylane_2022} and PyTorch~\cite{paszke_automatic_2017}. The energy consumption results estimated by CodeCarbon are multiplied by the Power Usage Effectiveness (PUE) for the cluster containing the VMs~\cite{gcp_pue}.

\subsection{QPU consumption estimates}

We used a 36-qubit trapped-ion quantum system, the IonQ Forte Enterprise\footnote{\url{https://ionq.com/quantum-systems/forte-enterprise}}, for actual quantum runs. Forte Enterprise features all-to-all connectivity between its qubits, which eliminates the need for SWAP-based routing overhead that constrains superconducting architectures and enables the direct execution of the entangling patterns in our ansatz without additional compilation overhead.

This system features electrical monitoring and logging, from which we can monitor power draw measurements, the total time elapsed, and thus the energy consumed over the duration of all submitted jobs. Monitoring is done for macroscopic groupings of system components, allowing us some additional capacity to track major sources of the overall power draw. The power draw measured by these components was logged at a frequency of 1 Hz for the duration of the study; when cross-referenced with the start and end times for each circuit executed, these measurements can be used to construct estimates for both the total energy consumption and average power draw observed over the span of the inference circuit job.

\subsection{Foundation Model, datasets, and tasks}

As~\cite{kim2025quantum} mirrors SetFit~\cite{tunstall_efficient_2022} in using a base model pre-trained on the
\texttt{paraphrase} data set\footnote{\url{https://huggingface.co/sentence-transformers/paraphrase-mpnet-base-v2}}, with fine-tuning performed on a custom version of the Stanford Sentiment Treebank (SST2)~\cite{socher_recursive_2013} binary classification dataset, we use the same pre-trained model and data here. Our goal is to emulate a ``real-world'' setting as much as possible, and this includes the data distribution. Our study only measures the cost of running inference circuits, as the comparison here between QPU and GPU is much more straightforward than it is for training, where costs depend additionally on the complexity of the optimizer chosen, as well as the method for computing quantum gradients~\cite{li_hybrid_2017} if using a gradient-based optimizer. Nonetheless, because inference may be considered to be part of an iterative training process, some of the high-level conclusions we draw can also apply to training.

\subsection{Energy Study Configuration}
\label{sec:hyperparameters}

With the exception of qubit count $Q$, the hyperparameters used for the energy consumption study are all fixed to those values shown in Table~\ref{tab:hyperparam_letters}. These trivially satisfy the $o(Q^2)$ scaling described in Table~\ref{tab:energy_consumption_assumption} of Appendix~\ref{sec:energy_consumption_cmp}, as the bounded circuit depth ensures the number of single and two-qubit gates grows linearly with respect to $Q$. For this study, we evaluated ten different model sizes defined by setting $Q$ to all even number values (inclusive) between 10 and 28. As discussed further in Section \ref{sec:accuracy_config}, since all training was done using classical simulators, trained weights were not produced for most of the data points discussed here. As these tests were only used to assess energy consumption, and not accuracy, all circuits were run using randomized weight initializations, and the measurement outcomes were discarded. Moreover, while these experiments do measure the effect of debiasing on energy consumption, we do not consider the effects of any postprocessing here. These circuits were run on 25 of the 250 latent vector embedding samples used in Section \ref{sec:accuracy_config} for the accuracy testing.

\begin{table}[hbt]
    \centering
    \begin{tabular}{c c c}
\specialrule{.1em}{.05em}{.05em}
        \textbf{Hyperparameter} & \textbf{Symbol} & \textbf{Value} \\ [0.5ex]
\specialrule{.05em}{.05em}{.05em}
        \underline{Q}ubits & $Q$ & $Q$ \\
        Number of \underline{E}ncoders & $E$ & $1$ \\
        \underline{R}e-upload number & $R$ & $4$ \\
        Number of \underline{M}ain blocks & $M$ & $2$ \\
        \underline{N}umber of re-uploading blocks & $N$ & $1$ \\
        \underline{B}atch Size & $B$ & $16$\\
        Number of \underline{S}hots & $S$ & $600$ \\ [1ex]
\specialrule{.1em}{.05em}{.05em}
    \end{tabular}
    \caption{Hyperparameters for the energy consumption study. All values are fixed except the qubit count $Q$.}
    \label{tab:hyperparam_letters}
\end{table}

\subsection{Accuracy Study Configuration}
\label{sec:accuracy_config}

Separately from the energy study, we conduct accuracy experiments to validate that the QPU can produce classification results comparable to noiseless simulation. This study uses the same model architecture, dataset, and debiasing pipeline described above in section \ref{sec:hyperparameters}, differing only in shot count allocation and evaluation scope.

All quantum models in this study were trained in simulation using a gate noise model built to emulate the Forte Enterprise system; the trained parameters were then deployed on the QPU for inference without retraining. We evaluate the same $N = 250$ test samples from the SST2 test split at each of five qubit counts: 10, 12, 14, 16, and 18. Using identical samples across all qubit counts ensures that any differences in accuracy reflect hardware and model scaling rather than data variation. As qubit count increases, so does the two-qubit gate count (from 35 at 10 qubits to 63 at 18 qubits, see Table~\ref{tab:accuracy_config}), and each additional entangling gate contributes a cumulative depolarizing error that broadens the output distribution and reduces the probability mass on signal bitstrings. Estimating $\langle Z \rangle$ to a given precision from this noisier distribution requires proportionally more measurement samples. Accordingly, we scale the total shot count with qubit count. The configurations are summarized in Table~\ref{tab:accuracy_config}.

\begin{table}[hbt]
    \centering
    \begin{tabular}{c c c c}
\specialrule{.1em}{.05em}{.05em}
        \textbf{Qubits} & \textbf{Total shots} & \textbf{Shots/variant} & \textbf{2Q gates} \\ [0.5ex]
\specialrule{.05em}{.05em}{.05em}
        10 & 500 & 20 & 35 \\
        12 & 1{,}000 & 40 & 42 \\
        14 & 2{,}000 & 80 & 49 \\
        16 & 5{,}000 & 200 & 56 \\
        18 & 20{,}000 & 800 & 63 \\ [1ex]
\specialrule{.1em}{.05em}{.05em}
    \end{tabular}
    \caption{Accuracy study configuration. Total shot counts are scaled with qubit count to compensate for the higher cumulative gate error in deeper circuits. Shots per variant equals total shots divided by 25 transpiled variants.}
    \label{tab:accuracy_config}
\end{table}

We compare QPU results against three baselines: (1) noiseless statevector simulation using the same trained model parameters, (2) noisy simulation incorporating the gate noise model used during training, and (3) classical machine learning baselines (support vector classifier and logistic regression) trained and evaluated on the same SetFit embedding vectors and SST2 data split. The noiseless simulation represents the theoretical upper bound for the quantum model, while the noisy simulation reflects the accuracy the model was optimized to achieve under realistic noise conditions during training.

We note that these shot counts differ substantially from the 600 shots used in the energy study (Section~\ref{sec:hyperparameters}). The energy study uses a fixed, modest shot count to enable fair scaling comparisons across qubit counts, while the accuracy study allocates sufficient shots to characterize the best achievable QPU performance at each scale.

\section{Analysis and Results}
\label{sec:results}

\subsection{Energy Consumption Scaling}
\label{sec:energy_results}

\begin{figure}[hbt]
    \centering\includegraphics[width=0.45\textwidth]{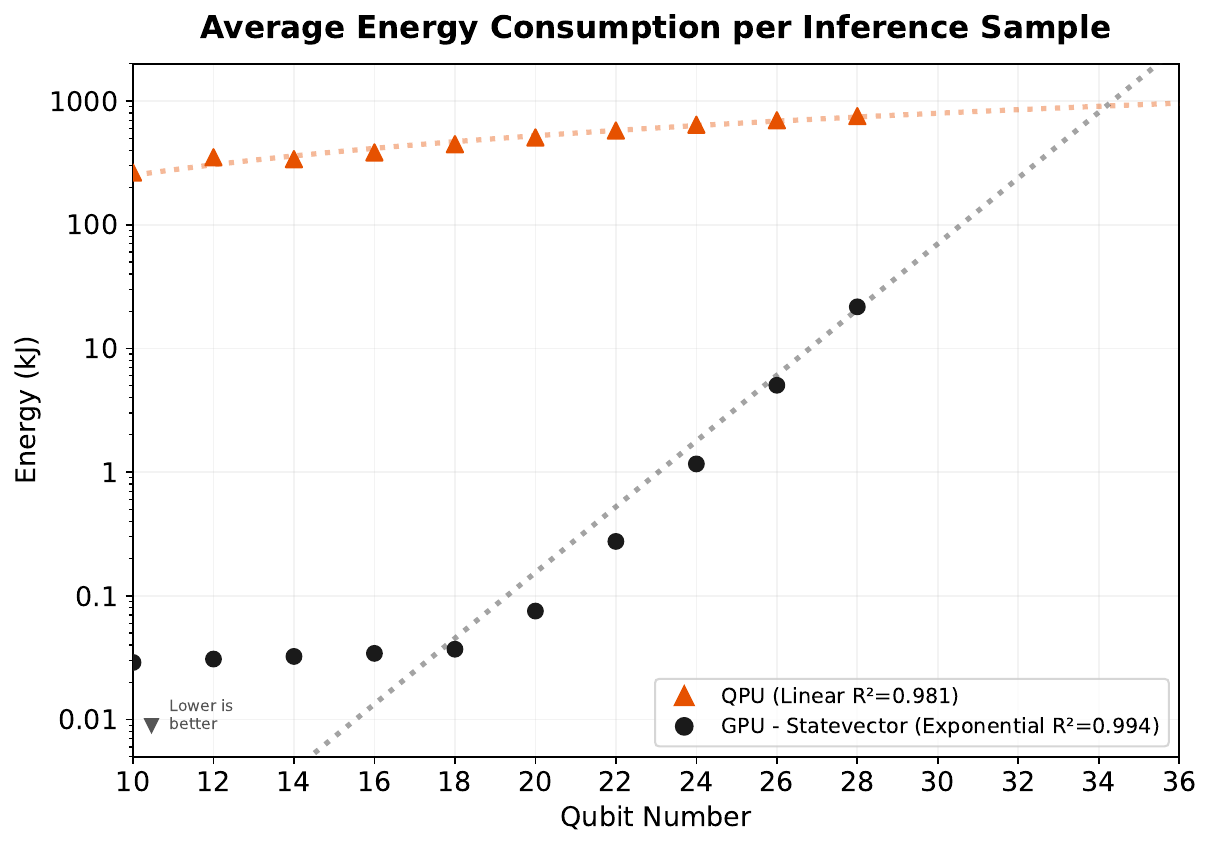}
    \caption{Average energy consumption, shown on a semilog plot, for 600-shot debiased inference circuits on QPU (IonQ Forte Enterprise) vs GPU (NVIDIA L4). Least squares fits strongly indicate linear and exponential scaling for QPU and GPU runs, respectively, and indicate break-even of energy-to-solution at around 34 qubits. The exact crossover point is specific to choices of hardware, circuits, and hyperparameters.}
    \label{fig:energy_consumption_estimate}
\end{figure}

Each latent vector embedding from the test set constitutes its own circuit job, which is then further split into 25 separate circuit executions by the debiasing procedure. Aggregating the power draw readings for all these executions allows us to construct the estimated energy consumption of each submitted job. We have averaged these energy estimates over each qubit number and shown them on a semilog plot in Figure~\ref{fig:energy_consumption_estimate}. Likewise, we also depict the estimate energy consumption of the GPU-accelerated statevector simulation runs, as measured by CodeCarbon on a single NVIDIA L4. These figures are averaged over 10 latent vector embedding calculations performed sequentially.

Trend lines are given based on a linear fit for the QPU runs and an exponential fit for the GPU runs.
At $R^2$ value of $0.994$, the exponential trend is a reasonable fit to the GPU data, matching the estimate formula reproduced in Appendix \ref{sec:energy_consumption_cmp} from~\cite{kim2025quantum}. This trend also matches natural intuition, as the number of computations required to execute a quantum circuit on a statevector simulation doubles with each additional qubit. Due to fixed computational overhead, the first three points corresponding to 10, 12 and 14 qubits visibly deviate from the exponential fit. This fixed overhead is overtaken at higher qubits and fitting the GPU data for data points corresponding to 16 qubits and beyond leads to an $R^2$ value of $0.999$. 

In contrast, the linear trend line fits to the QPU data with an $R^2$ value of $0.981$. This trend also matches the corresponding estimate formula from Appendix \ref{sec:energy_consumption_cmp}, which predicts linear energy scaling so long as the number of single- and two-qubit gates grows linearly with $Q$, as is the case with our ansatz. We do note that debiasing will meaningfully alter the overhead factors from what is presented in the appendix. This prediction depends on $P_{qpu}$, the average QPU power draw, behaving independently of $Q$, which we discuss further in Section \ref{sec:power_draw}. We do note that these data predict the crossover point, at which the QPU becomes energetically favorable, occurring around 34 qubits, but it is important to keep in mind that this extrapolation is specific to the choice of hardware (both QPU and GPU), choice of circuit, and hyperparameter configuration used in this study.

\subsection{Power Draw Analysis}
\label{sec:power_draw}

\begin{figure}[hbt]
    \centering
    \includegraphics[width=0.45\textwidth]{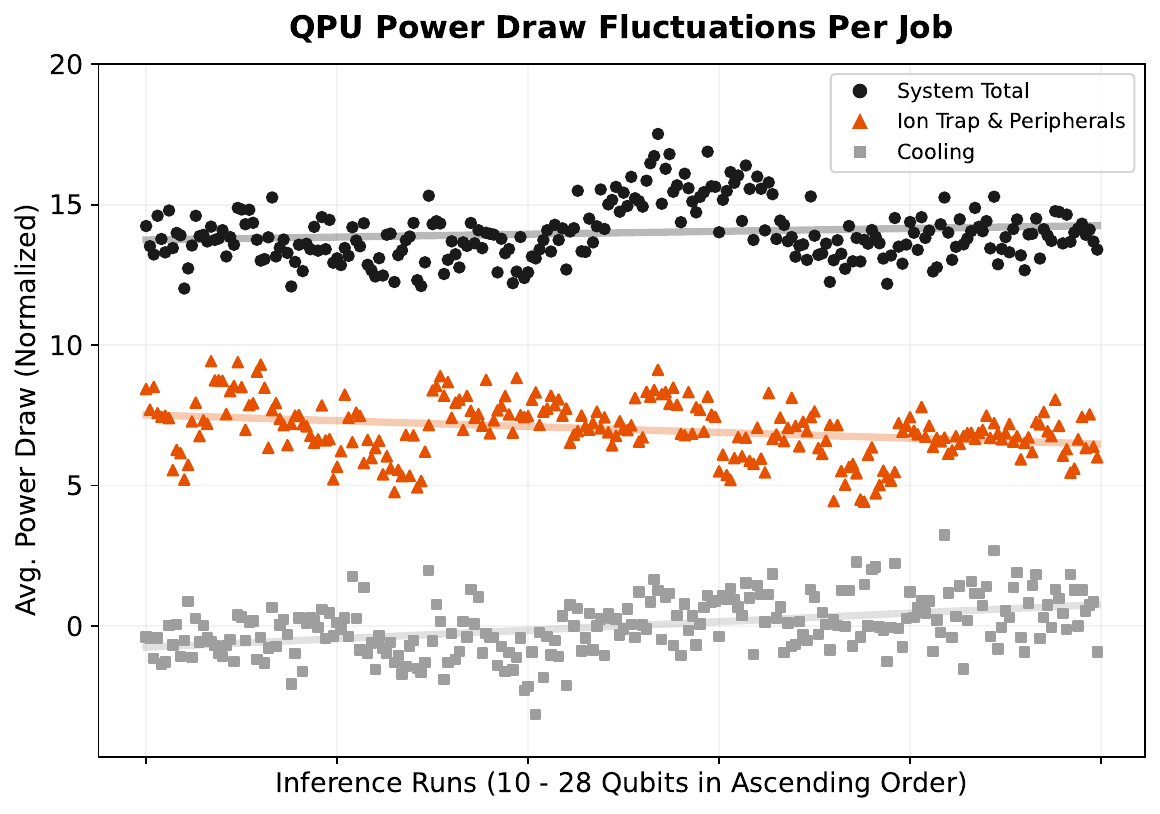}
    \caption{Per-job average power draw readings for one slate of inference tests, increasing from 10 to 28 qubits. Fluctuations have been normalized and shifted vertically for readability, and are shown for both the total system and the two macroscale components with greatest power draw.}
    \label{fig:power_draw_fluctuations}
\end{figure}

Consistent with known QPU energy consumption estimates from the literature (such as Equation~\ref{eq:energy_models}), which multiply the average QPU power draw by the execution time of the quantum circuit, we find that the QPU exhibits a relatively consistent power draw regardless of how large a circuit it operates on. This observation implies that time-to-solution is the only size-dependent factor of the total energy consumption. To corroborate our findings, we present Figure~\ref{fig:power_draw_fluctuations}, which shows data corresponding with the average power draw readings of each inference run used to produce the results in Section~\ref{sec:energy_results}. We give the data points for the entire QPU, alongside those of the ion trap, plus peripheral components, and the combined cryogenic and non-cryogenic cooling modules. These averages, presented in ascending qubit order from left to right, have been normalized and shifted vertically to highlight the qualitative relationships between the fluctuations of these three readings.

While mild upward trends in these fluctuations can be observed for the system as a whole and for the cooling components, they do not appear to indicate a strong relationship between the average power readings and the qubit number $Q$, for several reasons. The average power draw reading for the ion trap correlates negatively with the qubit number, and for all three trends, the relationship is far from monotonic. Our conclusion is that any discernible effect that circuit size could have on these power readings is completely washed out by dominant factors related to the general operating conditions of the QPU.



\subsection{QPU Accuracy Validation}
\label{sec:accuracy_results}

Having established the energy scaling behavior of the QPU, we now turn to the question of whether the quantum circuits produce accurate classification results when executed on hardware. This is a necessary complement to the energy analysis: favorable energy scaling is only meaningful if the QPU is performing useful computations.

\begin{figure}[hbt]
    \centering\includegraphics[width=0.45\textwidth]{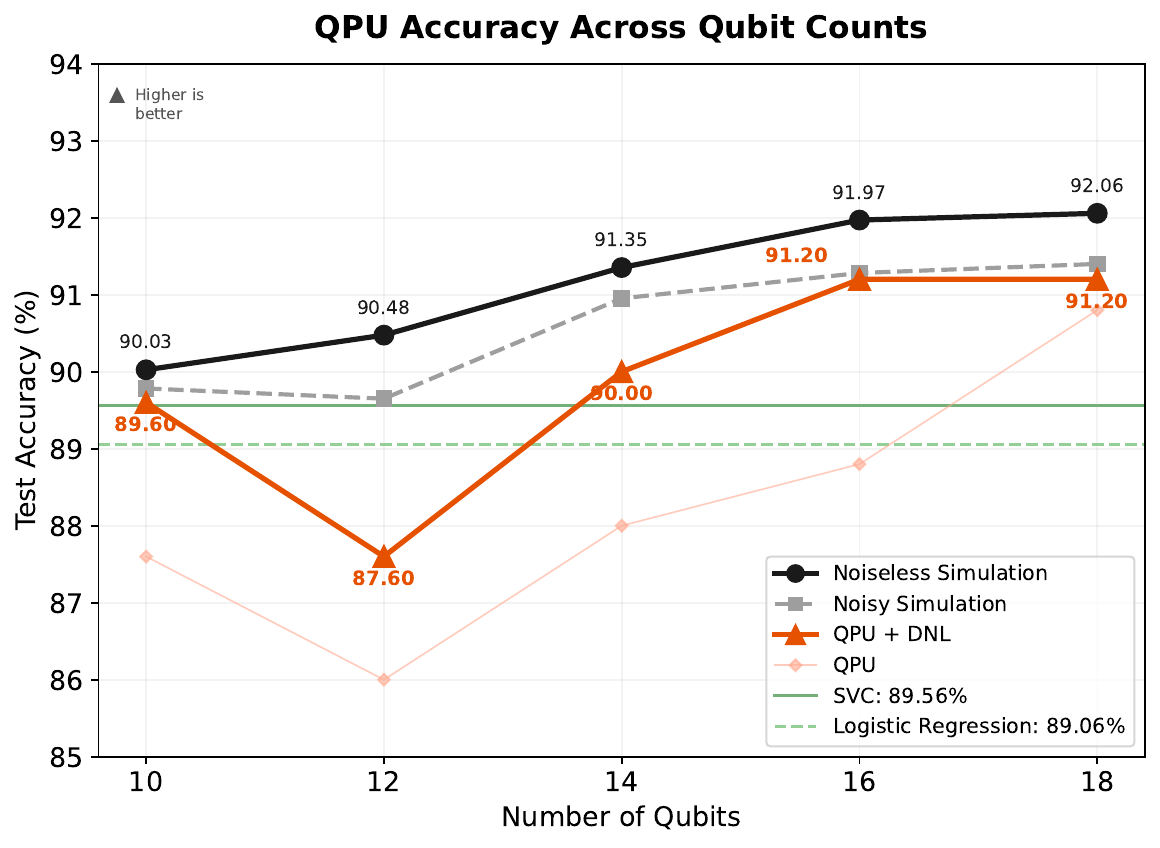}
    \caption{Test accuracy on SST2 binary classification as a function of qubit count. QPU+DNL results use debiasing and non-linear filtering for error mitigation. Also shown are noiseless and noisy (gate noise model) simulations, and classical baselines (SVC and logistic regression). The filtered QPU results approach simulation accuracy at scale, consistently exceeding classical baselines starting at 14 qubits. At 18 qubits, results show a reduction in the classical baseline error of $23.9$\% and $15.7$\% respectively when executing our model on a noiseless simulator and on a QPU with error mitigation.}
    \label{fig:qpu_accuracy}
\end{figure}

Figure~\ref{fig:qpu_accuracy} presents the classification accuracy across all five qubit counts for four conditions: noiseless simulation, noisy simulation (using the gate noise model from training), QPU with debiasing only, and QPU with the non-linear aggregation filter plus debiasing. Two classical baselines---a support vector classifier (SVC, 89.56\%) and logistic regression (89.06\%)---are shown as horizontal reference lines. Detailed results are given in Table~\ref{tab:accuracy_results}. The classification error improvement of the best quantum fine-tuned model over the best classical fine-tuned model (SVC) is around 23.9\% (Ideal) and 15.7\% (QPU). 

\begin{table}[hbt]
    \centering
    \begin{tabular}{c c c c c c}
\specialrule{.1em}{.05em}{.05em}
        \textbf{Q} & \textbf{Ideal} & \textbf{Noisy} & \textbf{QPU} & \textbf{+DNL} & \textbf{Gap} \\ [0.5ex]
\specialrule{.05em}{.05em}{.05em}
        10 & 90.03 & 89.78 & 87.60 & 89.60 & 0.43 \\
        12 & 90.48 & 89.65 & 86.00 & 87.60 & 2.88 \\
        14 & 91.35 & 90.95 & 88.00 & 90.00 & 1.35 \\
        16 & 91.97 & 91.28 & 88.80 & \textbf{91.20} & 0.77 \\
        18 & \textbf{92.06} & \textbf{91.40} & \textbf{90.80} & \textbf{91.20} & 0.86 \\ [1ex]
\specialrule{.1em}{.05em}{.05em}
    \end{tabular}
    \caption{Classification accuracy (\%) on 250 SST2 test samples. ``Ideal'' is noiseless simulation, ``Noisy'' is gate-noise-modeled simulation (training distribution), ``QPU'' is the debiased hardware result, ``+DNL'' denotes the addition of a non-linear aggregation filter on top of debiasing, and ``Gap'' is the difference between Ideal and +DNL. 18 qubits yielded the greatest accuracy across all regimes (denoted in bold).}
    \label{tab:accuracy_results}
\end{table}

Several observations emerge from these results.

\textit{The hardware--simulation gap closes at scale.} We report the QPU accuracy with debiasing only, 90.8\%, as the baseline hardware result at 18 qubits. We additionally report the filtered accuracy, 91.20\%, as the best calibrated hardware performance achieved in this study. This calibrated result is only 0.86 percentage points below the noiseless simulation ceiling of 92.06\%, retaining 99.1\% of ideal performance. At 16 qubits, the filtered result is 91.20\%, only 0.08 percentage points below the noisy training simulation (91.28\%).

\textit{QPU accuracy surpasses classical baselines.} From 14 qubits onward, the filtered QPU result consistently exceeds both SVC (89.56\%) and logistic regression (89.06\%) baselines, reaching 91.20\% at 16 and 18 qubits. Even without the filter, the raw QPU result at 18 qubits (90.80\%) surpasses both classical baselines. We note that with $N = 250$ samples, individual accuracy estimates have a standard error of $\approx 1.8$ percent; accordingly, the advantage over classical baselines at any single qubit count is modest in isolation, but the consistent trend across 14--18 qubits strengthens this finding.

\textit{Accuracy generally improves with qubit count.} Despite the increase in two-qubit gate count from 35 to 63 gates, test accuracy improves for both simulation and QPU conditions as qubit count increases. The 12-qubit configuration is a notable exception, where the filtered QPU accuracy (87.60\%) lags furthest behind ideal simulation (90.48\%). From 14 qubits onward, the overall trend indicates that the model is in a regime where the increased expressivity of larger circuits outweighs the additional noise, a favorable scaling property for future experiments on improved hardware.

\textit{The non-linear filter is most effective at intermediate shot regimes.} The benefit of the filter varies with the shot budget. At 16 qubits (5{,}000 total shots, 200 per variant), it improves accuracy by 2.4 percentage points (88.80\% to 91.20\%), shifting 10 borderline samples, 8 of which are corrected. At 18 qubits (20{,}000 total shots, 800 per variant), the improvement is only 0.4 percentage points (90.80\% to 91.20\%), as the higher shot count already provides sufficient statistics to suppress transpilation-level noise through simple averaging. This is consistent with the filter's role: it suppresses bitstrings that appear with anomalously high frequency in only a few variants, but when shot counts are high enough, the per-variant histograms contain less noise for the filter to remove.

We note that the shot counts used in this accuracy study (up to 20{,}000 total, or 800 per variant) substantially exceed those of the energy study (600 total). The energy comparison in Section~\ref{sec:energy_results} uses a fixed, moderate shot count to isolate the scaling relationship between qubit count and energy consumption. The accuracy study, by contrast, allocates sufficient shots to characterize the best achievable QPU performance and to validate that the quantum model is performing meaningful computation at each scale. The energy--accuracy trade-off --- that higher shot counts yield better accuracy but consume more energy --- is an important practical consideration, and we leave a detailed analysis of optimal shot allocation strategies to future work.

\section{Theoretical Energy Consumption Comparison with Matrix Product States}
\label{sec:mps_comparison}

As a fundamental classical baseline for comparing QPU performance, statevector simulation is worthwhile to consider because it performs the exact same calculations as the true quantum circuit. We are therefore focusing on benchmarking simulation cost with the present work and are considering, in a sense, a ``worst case'' classical approach. For future solution cost benchmarking, one would naturally want to consider more computationally favorable classical baselines (which may or may not exist, depending on the problem under consideration). Thus, we now devote some analysis toward a rough scaling comparison between executing NISQ circuits, of the form discussed in this paper, via QPU and via MPS simulation.

An MPS~\cite{schollwock2011density} is a way to represent multi-qubit wavefunctions as superpositions of matrix products, where each matrix corresponds with one basis state of a specific qubit: \begin{equation}
    |\psi\rangle = \sum_{q_1,\dots,q_Q} A_1^{q_1}\dots A_Q^{q_Q}|q_1,\dots,q_Q\rangle
    \label{eq:mps_defn}
\end{equation} The bond dimension $\chi$ of an MPS is the maximum dimension of any one of these constituent matrices. An MPS with bond dimension $2^{\frac Q 2}$ can represent any $Q$-qubit wavefunction, but to keep MPS calculations computationally tractable it is necessary to truncate $\chi$ in practice. This choice of $\chi$ determines the levels of inter-qubit entanglement that the MPS can model.

The dominant computational cost of simulating a quantum circuit with an MPS backend is $O(Q\chi^3)$~\cite{vidal2003efficient}. To effectively simulate a circuit with $D$ entangling blocks, which scales commensurately with circuit depth for hardware-efficient ans\"atze like those discussed in this work, an MPS must have a bond dimension of at least $2^D$\cite{vidal2003efficient, ran2020encoding}. As a result, restricting the circuit depth as we have done for ease of training does make possible the potential for efficient, linear simulation via an MPS. It is known that with global cost functions, circuit depths must be restricted to $O(1)$ scaling to avoid vanishing gradients, but with local cost functions like the one used in our architecture, effective circuit training can be done at $\log Q$ circuit depths~\cite{cerezo2021cost}. It follows for these logarithmic-depth circuits that the computational time, and therefore the energy cost, of effective MPS simulation scales as $O(Q^4)$.

Our experimental results in Section~\ref{sec:results} indicate that circuit execution time (including overhead) is the dominant variable factor contributing to QPU energy consumption. More specifically, a circuit of $Q$ qubits with depth $D$ would take $O(QD)$ time to execute, and the energy consumption will increase proportionally. Therefore, in contrast with the MPS backend, the energy cost of executing log-depth circuits on a QPU is $O(Q\log Q)$. Our results therefore indicate that while an MPS can simulate a log-depth NISQ circuit in polynomial time, it is more energetically favorable at scale to perform these calculations on quantum hardware. We conjecture that the crossover point at which QPUs become more energy-efficient compared to MPS simulations lies at a scale achievable by NISQ devices.

\section{Conclusion}
\label{sec:conclusion}

In this paper, we compared real measurements of the energy consumption required to run quantum algorithms for machine learning inference on both quantum and classical hardware. As an illustrative trial use case, we conducted these tests with an established hybrid architecture for fine-tuning pretrained classical language models using a parameterized quantum circuit as a fine-tuning head. Following established literature, we used a quantum head to fine-tune a text encoder foundation model into a binary classifier for performing sentiment analysis. Using the IonQ Forte Enterprise, we performed inference runs of this quantum head to analyze the energy consumption and accuracy of this model when run on quantum hardware.

Our collected data corroborates the intuitive understanding that for a constant-depth hardware-efficient circuit ansatz, execution time grows linearly with respect to qubit count. Moreover, we find no strong evidence for any significant relationship between the QPU's power draw and the size (width and depth) of the quantum circuit on which it operates: as a result, we find in our experiments that energy consumption scales linearly with qubit number, in contrast with the exponential computational time required to simulate these circuit executions on a GPU. Simultaneously, we demonstrate that with a few architectural modifications, this fine-tuning pipeline can operate on a QPU at levels of accuracy comparable with a noiseless simulator, and that the gap in performance between the two shrinks with additional qubits. Finally, we acknowledge that while the scope of our specific experiments still lies within the realm of efficient classical simulation via MPS, we discuss how our findings indicate that executing on quantum hardware is likely energetically favorable compared to MPS simulation at NISQ scales, even at circuit depths that are computationally classically tractable.

For future work, we consider that our analysis did not consider the energy consumed by training these models directly on quantum hardware. Additionally, large classical compute clusters like ORNL Frontier~\cite{noauthor_frontier_nodate} make it possible to directly compare the performance of QPU-executed quantum architectures with classical baselines beyond the scales analyzed in this work. An in-depth experimental analysis further comparing QPU performance with that of an MPS-simulated classical backend would be greatly beneficial for improving our understanding of the exact boundary between classical and quantum advantages in all its possible forms.

\appendices
\section{SST2 Dataset}
\label{sec:sst2_for_setfit}

As in~\cite{kim2025quantum}, we use a subset of the Stanford Sentiment Treebank (SST2) that was selected in developing SetFit\footnote{\url{https://huggingface.co/datasets/SetFit/sst2}}. This version of the dataset comprises $N=9,613$ sentences labeled as either positive ($1$) or negative ($0$) in sentiment. We create a balanced data set of $256$ samples from each label $\{1, 0\}$ for training (including validation, but this is not pertinent, as we do not use these samples for this study). The remaining ``test'' split of the data set contains $N-2\times 256 = 9101$ samples. Since we are concerned primarily with inference, we use ``test'' samples for evaluation.

\section{Energy consumption estimate}
\label{sec:energy_consumption_cmp}

We present, with mild modification accounting for overhead factors in the case of the QPU, formulae from~\cite{kim2025quantum} that estimate the total energy consumption needed to run a quantum circuit either on a QPU or as a classical GPU-accelerated statevector simulation, alongside reasonable estimates for the constant terms required to evaluate each expression. 

\begin{table}[hbt]
    \centering
    \begin{tabular}{c c}
\specialrule{.1em}{.05em}{.05em}
        QPU energy consumption ($P_{qpu}$) & $\sim5$ kW \\
        1Q gate time ($T_{sq}$) & $1.1\times10^{-4}$s \\
        2Q gate time ($T_{tq}$) & $9\times 10^{-4}$s \\
        Number of single-qubit gates ($SQ_Q$) & $o(Q^2)$ \\
        Number of two-qubit gates ($TQ_Q$) & $o(Q^2)$ \\
        Number of shots ($S$) & 600 \\
        Overhead per shot ($O_S$) & $1.5\times 10^{-3}$s\\
        Circuit Overhead ($O_C$) & $10$s\\
        L4 GPU max power draw ($P_{gpu}$) & 0.072 kW \\
        L4 GPU FP32 speed ($F_{gpu}$) & $3.03 \times 10^{13}$ FLOPS \\ [1ex]
\specialrule{.1em}{.05em}{.05em}
    \end{tabular}
    \caption{Specifications used for estimating energy consumption for both QPU and GPU. QPU coefficient values are rough ballpark estimates commensurate with the performance specifications of the IonQ Forte Enterprise.}
    \label{tab:energy_consumption_assumption}
\end{table}

$E_{qpu}$ and $E_{gpu}$ are the estimated energy consumption for QPU and GPU in kJ, respectively.
\begin{align}
\label{eq:energy_models}
    E_{qpu} &= \Big(\big(( SQ_Q \cdot T_{sq} + TQ_Q \cdot T_{tq}) + O_S\big) \cdot S + O_C\Big) P_{qpu} \nonumber \\
    E_{gpu} &= 2^{Q} \cdot \frac{(SQ_Q \cdot 2^{2} + TQ_Q \cdot 2^{3})}{F_{gpu}} \cdot P_{gpu}
\end{align}
where both the 1- and 2-qubit gate counts $SQ_Q$ and $TQ_Q$ are subscripted by the total qubit count $Q$, in order to denote that for reasonable choices of ansatz, including the ones discussed in this study, we should expect these quantities to scale with $Q$. GPU communication overhead across multiple devices is generally negligible when the statevector is distributed across multiple GPUs, as is the postprocessing overhead in the case that the circuit is split and knit. Thus it is reasonable to conservatively estimate the total energy consumption as if the entire job were performed by a single GPU, even in the event that a distributed multi-GPU setup were used~\cite{kim2025quantum}.

\bibliographystyle{ieeetr}
\bibliography{bibliography}

\end{document}